# Using the photo-induced $L_3$ resonance shift in Fe and Ni as time reference for ultrafast experiments at low flux soft X-ray sources

Somnath Jana[1, *], Shreyas Muralidhar[2], Johan Åkerman[2], Christian Schüßler-Langeheine[1], Niko Pontius[1]

[1]Helmholtz-Zentrum Berlin für Materialien und Energie GmbH, 12489 Berlin, Germany

[2]Department of Physics, University of Gothenburg, 412 96 Gothenburg, Sweden

Abstract:

We study the optical-pump induced ultrafast transient change of the X-ray absorption at the $L_3$ absorption resonances of the transition metals Ni and Fe in $Fe_{0.5}Ni_{0.5}$ alloy. We find the effect for both elements to occur simultaneously on a femtosecond timescale. This effect may hence be used as a handy cross-correlation scheme providing a time-zero reference for ultrafast optical-pump soft X-ray-probe measurement. The method benefits from a relatively simple experimental setup as the sample itself acts as time-reference tool. In particular, this technique works with low flux ultrafast soft X-ray sources. The measurements are compared to the cross-correlation method introduced in an earlier publication.

Introduction:

Availability of femtosecond ultrashort X-ray pulses [1] lead to exciting insights into a plethora ultrafast processes in solid state, molecular, or atomic physics [2–5]. Most common for time-resolved X-ray experiments is the combination of an ultrashort X-ray probing pulse with a synchronized laser pump pulse in the infrared, visible or ultraviolet range. Mandatory for properly conducted time-resolved experiments is a reliable reference for the time delay between the pump and the probe pulse, i.e. a time normal to which all measured data can be referred to. Ideally, such a time reference even provides an independent determination of time-zero, i.e. the time of simultaneous arrival of laser and X-ray pulse at the sample. The independent experimental determination of time-zero removes uncertainties of deriving a per se unknown dynamic response of the sample under investigation.

Different cross-correlation schemes have been developed for this purpose. Cross-correlation schemes usually provide an indirect measurement of time zero. In a sequential linear cross-correlation process, photons of one pulse create excited states in a suited sample that change its response, R(t). The other pulse probes this altered response as a function of delay. To derive time zero from this measurement with high reliability, the response R(t) has to be known precisely and involve timescales shorter than or similar to the experimental resolution. Finally, the probed response R(t) is experimentally blurred by the temporal duration of the exciting and probing pulse. Effectively, the experimentally probed response corresponds to the convolution of the time dependent response R(t) with the experimental temporal resolution allowing to experimentally deduce time zero from this measurement.

For ultrashort X-ray sources without any natural intrinsic synchronization, e.g. SASE free electron lasers, an experimental shot-by-shot analysis of the relative time relation of X-ray and laser pulses is crucial to achieve a sub 100 fs synchronization and according experimental time-resolution. To this end, several X-ray-optical cross-correlation schemes on a shot-to-shot basis have been put forward and are routinely used [6–10]. But even for ultrafast X-ray sources with an inherent, natural synchronization between X-ray and laser pulse (e.g. high-order harmonic generation [11,12], laser-driven plasma [9–11], or storage-ring based slicing sources [16,17]), an independent absolute time reference is generally essential to put the measured transients onto a proper time scale. This is particularly important when the dynamics of different atomic species or of separate subsystems probed through different observables shall be compared. So do, for example, the recent results of delays between the electronic and magnetic responses in metallic magnets [18–20] and delayed response in Ni demagnetization with respect to the Fe in FeNi alloy [21,22] measured at extreme ultraviolet energies emphasize the need for an independent time-zero reference at soft x-ray energies. Beyond these physical requirements, a regular referencing of the time scale during extended measurements may be essential for correction of slow experimental drifts of time-zero e.g. caused by thermal changes of the optical path lengths of the laser or X-ray branches.

---

[*] Author to whom correspondence should be addressed: sj.phys@gmail.com.

Several common cross-correlation schemes base on dynamic effects induced by high power X-ray pulses [7,9,23–28], leading to a transiently altered material response, which is then probed by the optical laser pulse. These are well suited for intense ultrashort X-ray sources like FELs but do not work for low X-ray fluxes like at HHG or slicing sources. In Ref. [29,30], a soft x-ray-optical cross-correlator scheme has been put forward capable of providing a defined time reference for low flux soft X-ray sources with tunable photon-energy. It utilizes the laser excited displacive coherent phonon oscillation (DCPO) in a Molybdenum-Silicon (Mo/Si) multilayer [31] as the pump induced response. The DCPO is probed through intensity variations of the first, second or third order X-ray Bragg diffraction peak of the Mo/Si layers super-structure. The phase of this coherent oscillation provides a defined time reference for the time-resolved laser-pump and X-ray probe experiment. The phase shift of the DCPO in MoSi, with respect to the effective time zero of the pump-probe experiment, however, was experimentally not determined. Therefore, this cross correlator so far only supplies a relative time reference.

In this report, we study another soft X-ray-optical cross-correlation scheme suited for low-flux X-ray sources. It is based on a pump-laser induced transient change of the optical $L_3$-resonance X-ray absorption spectrum (XAS) observed in different $3d$ transition-metal systems [3,32]. This transient response reflects the excitation of the $3d$ valence electron system occurring within a few femtoseconds after photoexcitation by the ultrashort laser pulse. We study this effect for Ni and Fe in a $Fe_{0.5}Ni_{0.5}$ alloy sample utilizing linearly polarized X-ray pulses. The transient absorption changes are compared to the phase determined from the measured DCPO cross-correlation [29], which is measured concurrently. We find that fitting the transient change of the X-ray absorption by modeling the laser induced electronic excitation and relaxation process allows an estimation of the time of photoexcitation i.e. time zero with a precision of a few tens of femtoseconds. The advantage of the proposed cross-correlation measurement is a relatively simple experimental setup and moreover, e.g. in magnetic dynamic studies, the sample itself can serve as cross correlator. The study concludes by using the transient X-ray absorption change to determine time zero with respect to the DCPO phase.

When an X-ray photon is absorbed at the $L_2$ and $L_3$ absorption resonances of Fe or Ni (or any other $3d$ transition metal ion), a $2p$ core electron is promoted into the $3d$ valence shell creating a $2p$ core hole state with $j = 1/2$ and $3/2$, respectively [33]. These optical transitions lead to characteristic element specific absorption resonances at defined energies as shown in the inset of Fig. 1(a) for Ni $L_3$ in FeNi alloy.

Resonant X-ray absorption in $3d$-transition metals can change immediately upon photo-excitation by an ultrashort laser pulse – even on a sub-fs scale - as recently shown in Ti [34] and Ni [35] in the EUV spectral range. In the soft X-ray spectral range, transient changes especially of the $L_3$ absorption resonance have been observed and studied [32,36–38]. As shown in the inset of Fig. 1(a) (at Ni $L_3$ edge), this transient modification is most pronounced at the low energy slope of the $L_3$ resonance, where it appears as a transient increase of absorption. Overall, the effect looks like a red-shift of the absorption resonance. The absorption at the low energy slope increases; afterwards it was found to partially recover within the subsequent picoseconds [32,36–38]. Similar ultrafast modifications of the $L_3$ absorption spectrum have been reported for Cobalt in a CoPd-multilayer [39] as a non-linear effect, when probing the sample absorption by single, intense ultrashort X-ray pulses.

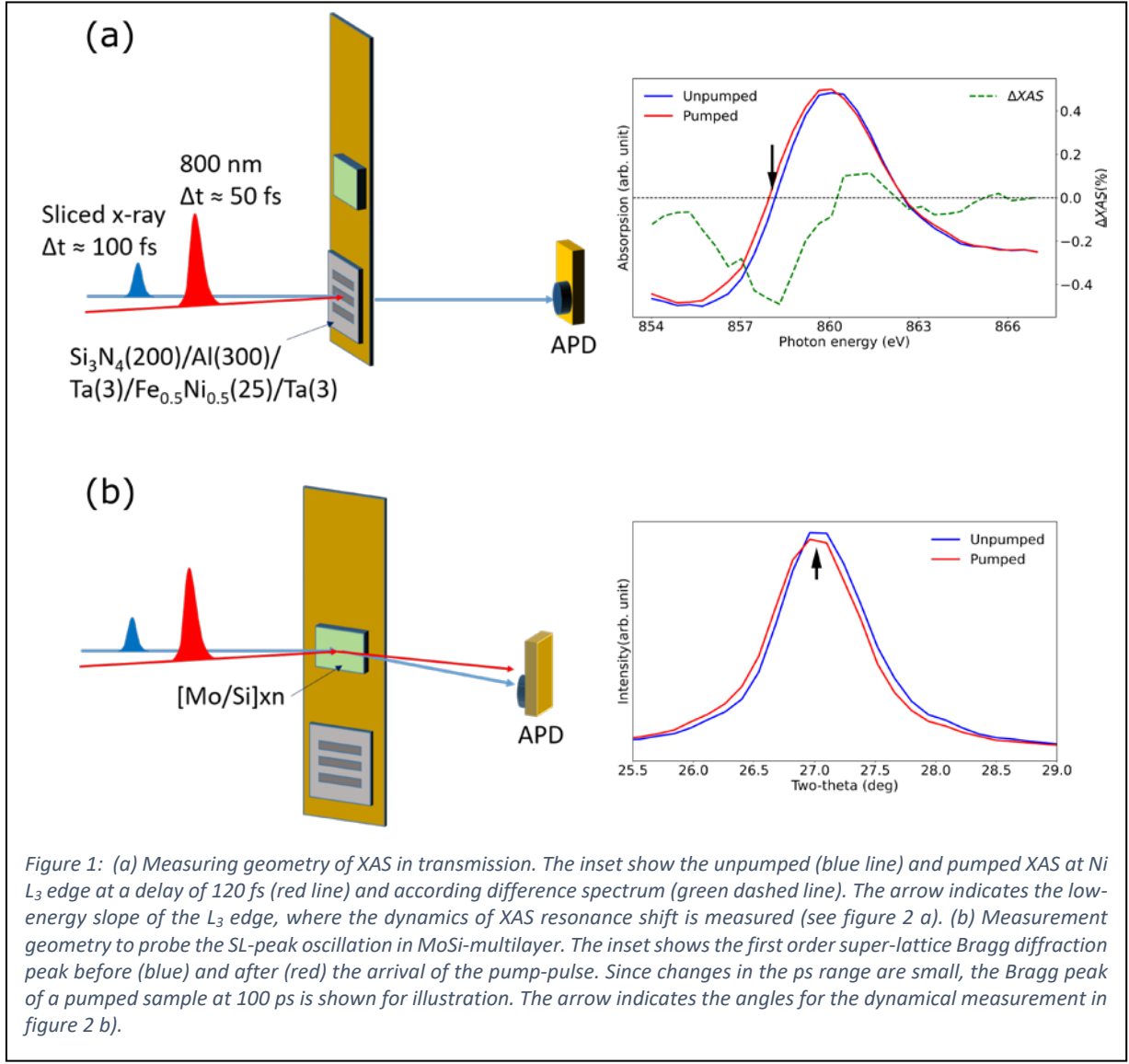

Figure 1: (a) Measuring geometry of XAS in transmission. The inset show the unpumped (blue line) and pumped XAS at Ni $L_3$ edge at a delay of 120 fs (red line) and according difference spectrum (green dashed line). The arrow indicates the low-energy slope of the $L_3$ edge, where the dynamics of XAS resonance shift is measured (see figure 2 a). (b) Measurement geometry to probe the SL-peak oscillation in MoSi-multilayer. The inset shows the first order super-lattice Bragg diffraction peak before (blue) and after (red) the arrival of the pump-pulse. Since changes in the ps range are small, the Bragg peak of a pumped sample at 100 ps is shown for illustration. The arrow indicates the angles for the dynamical measurement in figure 2 b).

These transient absorption changes have been assigned to pump-laser induced *3d* valence electron excitations, even though the descriptions of the process vary in their details. Refs. [32,36,37] relate the ultrafast $L_3$ modification to an increase of the *3d* electron localization resulting from a transient reduction of the nearest neighbor *3d-4sp* hybridization in Ni as a consequence of the electronic excitations. Similar spectral changes in the Ni $L_3$ absorption have been calculated by theory [40] and attributed to state blocking, i.e. a redistribution of 3d valence electrons from below to above the Fermi level through the optical excitation. The subsequent recovery of the absorption on a picosecond timescale has been assigned to the subsequent cooling and relaxation of the valence electron system through thermal coupling to the lattice.

Even if the exact physical mechanism behind the absorption changes at the *3d* metals $L_3$ resonances upon laser excitation is still under discussion, a common base of all interpretations is that it is associated with the pump-induced excitation of the 3d valence electron system. We hence would expect the magnitude of the effect to depend on the number of valence electrons excited from below to above the Fermi edge. The maximum number of excited electrons is reached, when the photo-absorbed energy has spread in the valence electron system through inelastic electron-electron scattering cascades. This occurs within a few tens of femtoseconds after photoexcitation [41–43] and leads to a Fermi-Dirac-like hot electron distribution. Subsequently, the number of excited electrons reduces when the electron system cools down, leading to the recovery effect seen in the $L_3$ absorption change.

Experiment:

As in Ref. [29] the experiments were performed at the BESSY II Femtoslicing beamline (UE56/1-ZPM) [44]. Vertically polarized x-rays were used to measure both, the XAS in transmission and the super-lattice Bragg peak in reflection (see Fig. 1). We used a $Fe_{0.5}Ni_{0.5}$ alloy thin film to measure the time dependent $L_3$ resonance XAS and a $[Mo(1.86nm)/Si(2.07nm)]_{40}$ multilayer to probe the oscillation of the first order super-lattice peak (SL1) at both, the Fe and the Ni $L_3$ edge energies. The $Fe_{0.5}Ni_{0.5}$ (25 nm) sample has been deposited on a $Ta(3\ nm)/Al(0.3\ \mu m)/Si_3N_4(0.2\ \mu m)$ substrate, where the Al layer acts as a heat sink and the 3 nm Ta as a seed layer. Another 3 nm Ta capping protects the sample from oxidation when exposed to air. An infrared laser of 800 nm wavelength, 3 kHz repetition rate and approx. 50 fs pulse duration was used to excite the samples. The pump laser beam cross section was with 0.35 x 0.25 mm$^2$ larger than the corresponding X-ray probe size of 0.15 x 0.04 mm$^2$. The XAS measurements were performed at 45° incident angle, to increase the effective thickness and with that the absorption of the $Ni_{0.5}Fe_{0.5}$ sample.

For the detection of the super lattice Bragg peaks the sample surface relative incident angle for the measurement was 13.5° at the Fe edge energy (~707 eV) and 11.2° at the Ni edge energy (~858 eV) (compare Fig. 1). The pump fluence was set to about 80% of the damage threshold for both the samples in order to maximize the pump-induced change without risking sample degradation. The data acquisition was performed by alternatingly measuring the X-ray absorption signal with and without pump pulse (referred to in the following as pumped and unpumped signal, respectively) on a shot-by-shot basis. Here the slicing source was running at 6 kHz and the pump laser at 3 kHz, allowing eliminating intensity fluctuations slower than the repetition rate [44]. To further eliminate slow drifts of time zero, our measurement protocol implemented alternating data collection between the alloy and the Mo/Si multilayer sample on an hourly time scale. The detail of the correction procedure are given in the appendix.

Fig. 2 (a) and (c) show the transient relative absorption change measured at the low energy $L_3$ resonance slope of Fe and Ni (compare Fig. 1 (a), inset for Ni), respectively. The symbols (circles and squares for Fe and Ni, respectively) represent the experimental data, the dashed lines are the least squares fits to the experimental data (see below). The measured transients are obtained from the measured data by calculating the relative change of the pumped XAS signal with respect to the unpumped. Both XAS transients show a sharp absorption increase up to about 0.4%, which peaks at about 100 fs delay. A slower gradual drop follows the initial quick rise. While the Ni transient levels off at about 0.25 % after one ps, the Fe transient changes sign after 0.5 ps and decreases to -0.15 %.

We model the XAS cross-correlation function, $C_{XAS}(t)$, as the convolution of the experimental resolution, $P(t)$, and the transient response from the sample $R_{XAS}(t)$:

$$C_{XAS}(t) = P(t) * R_{XAS}(t). \tag{1}$$

$P(t)$ is a normalized Gaussian function with FWHM = 130 fs [37] modeling the experimental temporal resolution; $R_{XAS}(t)$ is assumed as:

$$R_{XAS}(t) = H(t-t_0) \times \left\{a - b \times \left(1 - \exp\left(-\frac{t-t_0}{t_r}\right)\right)\right\}.$$

Here, the first term is a Heaviside step function scaled by *a*. This approach assumes the photo-induced XAS change to be quasi instantaneous on the time scale of the experimental temporal resolution. The second term takes care of the relaxation of the signal modeled through a single exponential decay. The fits are shown in Fig. 2 (a) and (c).

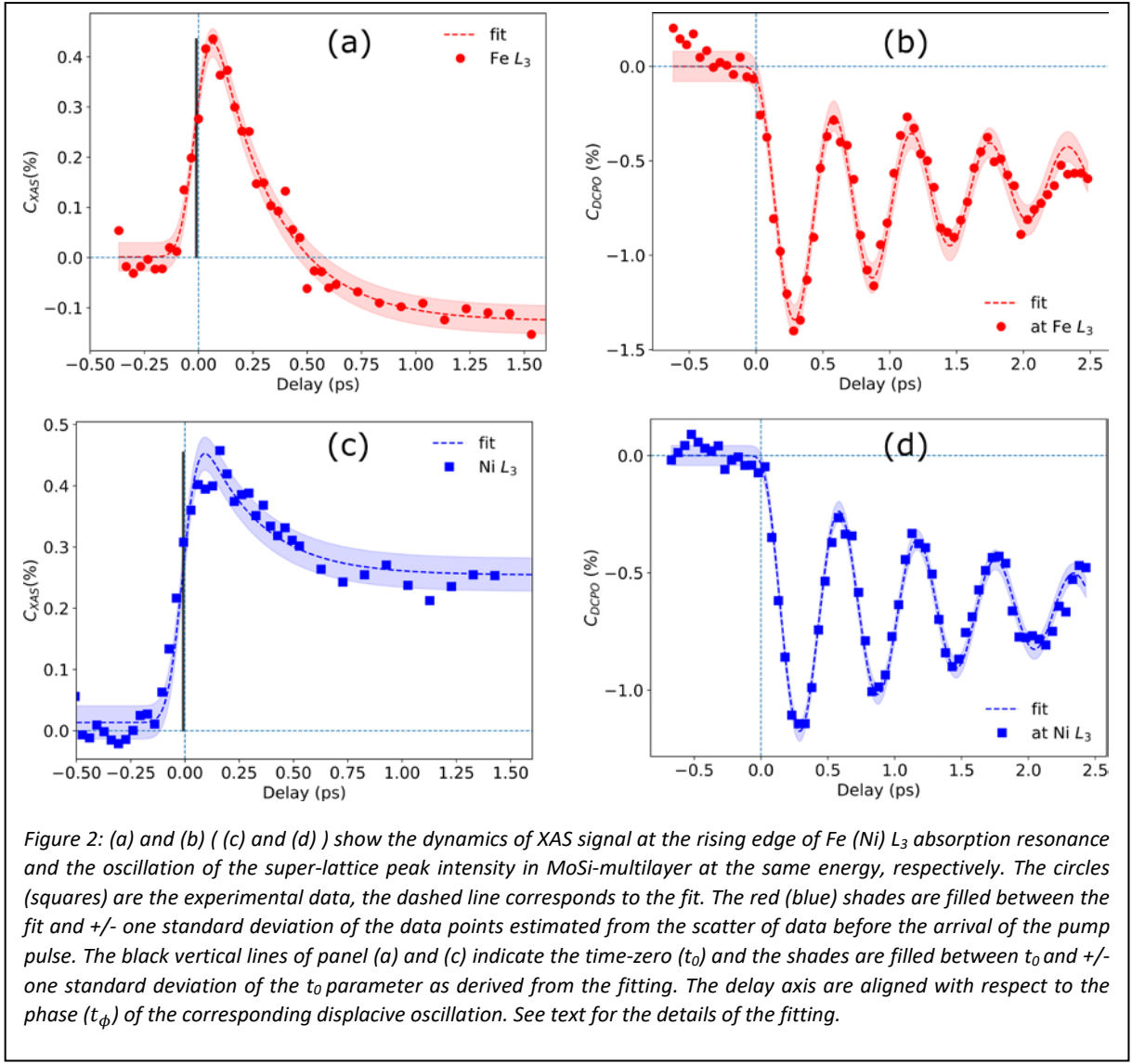

Figure 2: (a) and (b) ( (c) and (d) ) show the dynamics of XAS signal at the rising edge of Fe (Ni) $L_3$ absorption resonance and the oscillation of the super-lattice peak intensity in MoSi-multilayer at the same energy, respectively. The circles (squares) are the experimental data, the dashed line corresponds to the fit. The red (blue) shades are filled between the fit and +/- one standard deviation of the data points estimated from the scatter of data before the arrival of the pump pulse. The black vertical lines of panel (a) and (c) indicate the time-zero ($t_0$) and the shades are filled between $t_0$ and +/- one standard deviation of the $t_0$ parameter as derived from the fitting. The delay axis are aligned with respect to the phase ($t_\phi$) of the corresponding displacive oscillation. See text for the details of the fitting.

Fig. 2 (b) and (d) display the measured DCPO of the Mo/Si-multilayer each probed on the corresponding SL1-Bragg peak at identical photon energies as used in (a) and (c), respectively. Both traces show the first three DCPO oscillation periods as known from Ref. [29]. As described in the experimental section, the XAS and the DCPO transients of each photon energy has been measured in repeated alternation, which means that they are plotted on identical delay scale (see appendix).

The transient response for the DCPO cross-correlation of the MoSi sample is described correspondingly by

$$C_{DCPO}(t) = P(t) * R_{DCPO}(t). \qquad (2)$$

Here, the response function $R_{DCPO}(t)$ of the DCPO is assumed as

$$R_{DCPO}(t) = 0, \qquad for\ t \leq t_0,$$

$$R_{DCPO}(t) = A \times (O(t) - 1) + B \times (t - t_0), for\ t > t_0.$$

Here O(t) describes the damped cosine like DCPO by

$$O(t) = \cos(2\pi(t - t_\phi)/p) \times \exp(-(t - t_\phi)/T_r),$$

with $T_r$ the damping time of the oscillation amplitude, $p$ the oscillation period, and $t_\phi$ is the phase offset of the cosine function with respect to time zero ($t_0$). The latter entails the delay of the oscillation onset, which is related to the electron-phonon interaction time in Mo. The constant $A$ represents the amplitude of the displacement leading to an intensity drop of the SL peak intensity upon laser excitation. The linear term $B \times (t - t_0)$ models a slow linear change of the oscillation offset with time [45]. The fit results of $C_{DCPO}(t)$ to the experimental data are shown as dashed lines in Fig. 2 (b) and (d).

The most relevant individual fitting parameters for the experimental data are summarized in Table 1 and Table 2 with their one-sigma uncertainties. The mechanism for DCPO does not depend on the probe X-ray photon energy and should yield apart from statistical errors the same time dependence for the Ni and Fe photon energies. We therefore relate the time axes of both Fe and Ni transient XAS change to the DCPO measurement tentatively using $t_\phi = 0$ as time-zero reference.

*Table 1: Results obtained from the fitting of the oscillation of SL1 peak intensity in Mo/Si sample at Fe and Ni $L_3$ edges presented in Fig. 2(b), (d). The errors correspond to the one sigma values.*

|  | 707 eV (Fe) | 849 eV (Ni) |
|---|---|---|
| $A$ (%) | 0.8 ± 0.02 | 0.67 ± 0.02 |
| $p$ (fs) | 582 ± 6 | 585 ± 6 |
| $t_\phi$ (fs) | 0 ± 5 | 0 ± 4 |

*Table 2: $t_0$ obtained from the fitting of the $C_{XAS}$ data at Fe and Ni $L_3$ edges presented in Fig. 2(a) and (c). The errors correspond to the one sigma values.*

|  | Fe | Ni |
|---|---|---|
| $t_0$ (fs) | - 10 ± 5 | -7 ± 6 |

The obtained onset ($t_0$) from the fitting of the XAS transients reveals almost simultaneous response for Ni and Fe ($\Delta t_0$ = -3 ± 8 fs; compare Table 2) within our experimental error bar. The transient XAS change for Fe and Ni can hence be used as time reference for ultrafast dynamics experiments with a precision of approx. ten fs using the rather simple fit model presented above. The different recovery behavior of the two ions with the sign change found only in Fe suggest that a more elaborate model may further improve the precision as would a better statistics of the data. Furthermore both the XAS transients start almost simultaneously with a DCPO oscillation with phase $t_\phi = 0$, which hints towards a quasi-instantaneous response of the DCPO upon laser excitation. Relative to $t_\phi = 0$, the obtained $t_0$ from the XAS transients are -10 ± 7 fs and -7 ± 6 fs (see Table 1 and Table 2) for the Fe and Ni $L_3$ edges, respectively, which implies an onset of the DCPO delayed by 9 ± 9 fs.

A $t_\phi$ value close to zero appears reasonable in particular since we performed the DCPO measurement at a rather high pump fluence. While in general the phase lag can be fluence dependent, it has been shown in reference [45] that the phase lag converges to zero for the high fluence in the case of SrRuO$_3$/SrTiO$_3$ multilayer system.

We notice that the fit to the Ni XAS rising slope does not describe the experimental data as well as for the Fe. For Ni we get a better agreement assuming a temporal resolution of 220 fs instead of 130 fs. That a worse temporal resolution resulted from an incomplete drift correction for Ni appears rather unlikely (compare appendix). Instead, this observation could possibly hint to a built-up of the Ni XAS change taking a finite amount of time in the order of a few tens of fs instead of occurring almost instantaneously like in Fe.

In the present study we utilized linearly polarized X-ray pulses. The XAS transient change will be detectable with circular polarized X-rays as well. Then, however, a separation of the electronic response from magnetic dynamics is required, achieved, e.g., by averaging the signals measured for opposite circular helicities or magnetic field orientations, respectively.

In summary, we characterized a new cross-correlation experiment for an ultrafast optical-pump soft X-ray-probe experiment suited for low intensity X-ray sources. The technique is bases on probing the transient absorption change at the $L_3$ resonances of the 3d transition metals Ni and Fe. Time zero is estimated from the experimental traces by associating the transients with the quasi instantaneous excitation of the *3d* valence electron system by the optical pump laser and the following relaxation. Within the measurements of this study we are able to determine time zero by the proposed XAS cross-correlation method within an experimental error of 7 fs.

### Acknowledgement:

This work was funded by the German Research Foundation (DFG) through TRR227 (project A03).

### Appendix:

We have collected the data on the $Fe_{0.5}Ni_{0.5}$ alloy and the Mo/Si sample alternatively on an hourly time scale to correct for slow drifts of time-zero throughout the entire experiment performed at each resonance energy. Fig. 3 (a) and (b) show the variation of the time-zero versus the measurement time obtained at Fe and Ni *$L_3$* energies, respectively. In this context, time-zero ($t_0$) refers to the onset of the DCPO of the Mo/Si multilayer assuming the phase offset to be zero (see the main manuscript for the detail of the fitting function). As any individual DCPO scan is not suitable for refinement of all the parameters due to large noise, the average of all DCPO scans is fitted to obtain first guess values of different parameters, which are then used as fixed parameters to refine the onset ($t_0$) of any individual DCPO scan. Such obtained $t_0$ is then used to shift the delay-axis of the individual DCPO scan in order to calculate the average of all DCPO scans for the next iteration. The whole process is repeated interactively until the convergence of each $t_0$ parameter to within ± 1 fs is achieved. Fig. 1 (a) and (b) show the drift of the $t_0$ with time. A smoothed variation of the drift (dashed curve) is obtained by applying a Savitzky-Golay filter (with size 7 and order 3) three times consecutively on the obtained time zero variation (blue squares). The delay-axes of the transient XAS scans measured between the DCPO measurements are corrected by assigning their times to the smoothed curve.

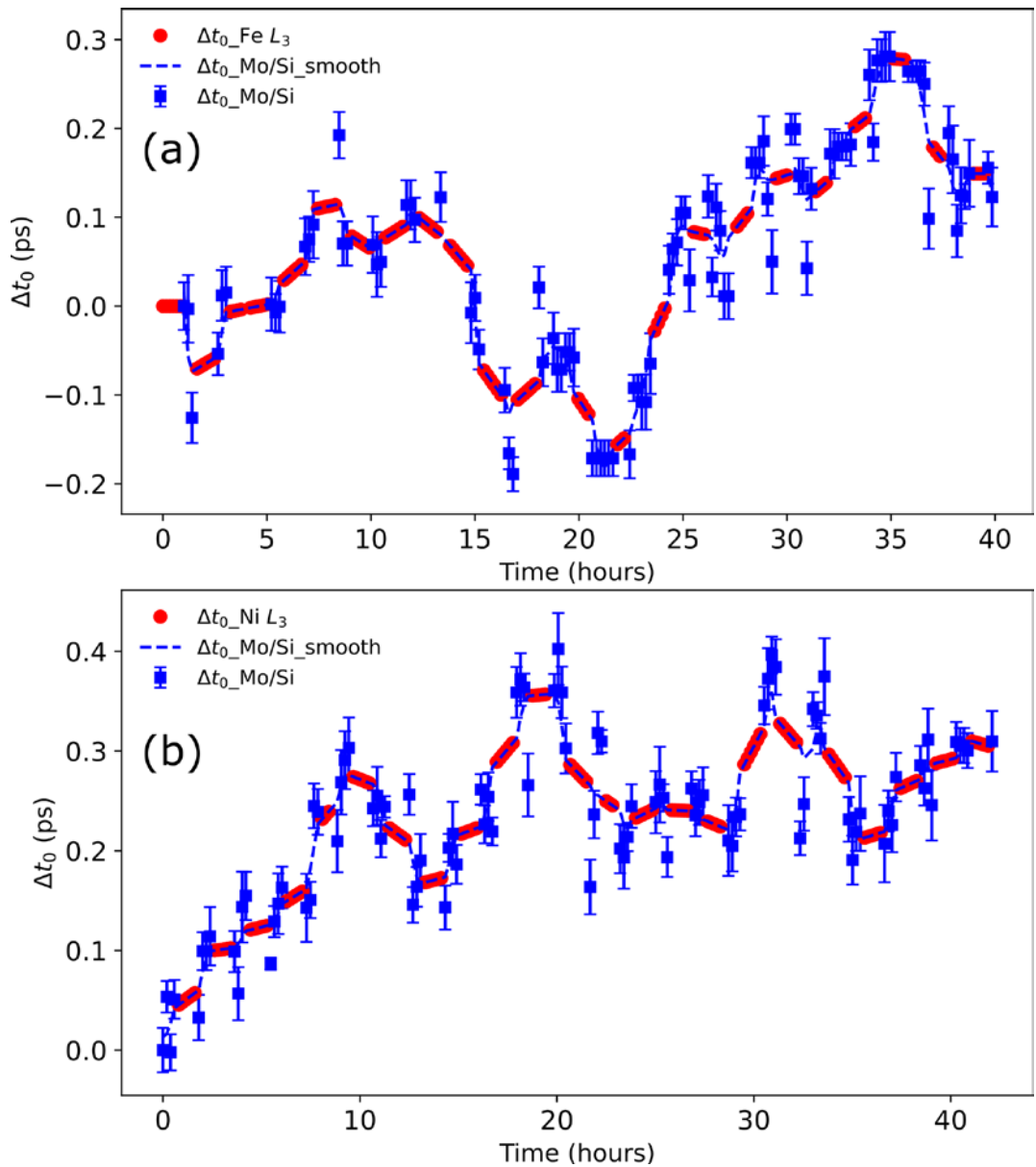

*Figure 3: Panel (a) and (b) show the variation of time-zero as a function of time at the Fe and Ni $L_3$ edges, respectively. The blue squares are obtained from the fitting of each individual DCPO scan. The dashed-line is obtained by applying a Savitzky-Golay filter as described in the text. The error bars correspond to the one standard deviation of the time-zero parameter of the fit-function when all other parameters are kept fixed (see text for the detail). The red circles are the time-zero-offset for the individual delay scans measured in the FeNi-alloy, obtained by assigning their times to the smoothed curve (dashed-line).*

### Data Availability:

Data available on request from the authors. (Note: We are aiming to publish the raw data to public repository that issues datasets with DOIs before this manuscript is published and would update the data availability information.)

References:

[1] R. Schoenlein, T. Elsaesser, K. Holldack, Z. Huang, H. Kapteyn, M. Murnane, and M. Woerner, *Recent Advances in Ultrafast X-Ray Sources*, Philos. Trans. R. Soc. A Math. Phys. Eng. Sci. **377**, 20180384 (2019).

[2] P. Wernet, K. Kunnus, I. Josefsson, I. Rajkovic, W. Quevedo, M. Beye, S. Schreck, S. Grübel, M. Scholz, D. Nordlund, W. Zhang, R. W. Hartsock, W. F. Schlotter, J. J. Turner, B. Kennedy, F. Hennies, F. M. F. De Groot, K. J. Gaffney, S. Techert, M. Odelius, and A. Föhlisch, *Orbital-Specific Mapping of the Ligand Exchange Dynamics of Fe(CO)5 in Solution*, Nature **520**, 78 (2015).

[3] C. Stamm, T. Kachel, N. Pontius, R. Mitzner, T. Quast, K. Holldack, S. Khan, C. Lupulescu, E. F. Aziz, M. Wietstruk, H. A. Dürr, and W. Eberhardt, *Femtosecond Modification of Electron Localization and Transfer of Angular Momentum in Nickel*, Nat. Mater. **6**, 740 (2007).

[4] C. Rose-Petruck, R. Jimenez, T. Guo, A. Cavalleri, C. W. Siders, F. Ráksi, J. A. Squier, B. C. Walker, K. R. Wilson, and C. P. J. Barty, *Picosecond-Milliangstrom Lattice Dynamics Measured by Ultrafast X-Ray Diffraction*, Nature **398**, 310 (1999).

[5] A. Cavalleri, C. Tóth, C. W. Siders, J. A. Squier, F. Ráksi, P. Forget, and J. C. Kieffer, *Femtosecond Structural Dynamics in Vo2 during an Ultrafast Solid-Solid Phase Transition*, Phys. Rev. Lett. **87**, 237401 (2001).

[6] B. Krässig, R. W. Dunford, E. P. Kanter, E. C. Landahl, S. H. Southworth, and L. Young, *A Simple Cross-Correlation Technique between Infrared and Hard x-Ray Pulses*, Appl. Phys. Lett. **94**, (2009).

[7] M. R. Bionta, H. T. Lemke, J. P. Cryan, J. M. Glownia, C. Bostedt, M. Cammarata, J.-C. Castagna, Y. Ding, D. M. Fritz, A. R. Fry, J. Krzywinski, M. Messerschmidt, S. Schorb, M. L. Swiggers, and R. N. Coffee, *Spectral Encoding of X-Ray/Optical Relative Delay*, Opt. Express **19**, 21855 (2011).

[8] S. Düsterer, P. Radcliffe, C. Bostedt, J. Bozek, A. L. Cavalieri, R. Coffee, J. T. Costello, D. Cubaynes, L. F. DiMauro, Y. Ding, G. Doumy, F. Grüner, W. Helml, W. Schweinberger, R. Kienberger, A. R. Maier, M. Messerschmidt, V. Richardson, C. Roedig, T. Tschentscher, and M. Meyer, *Femtosecond X-Ray Pulse Length Characterization at the Linac Coherent Light Source Free-Electron Laser*, New J. Phys. **13**, (2011).

[9] M. Beye, O. Krupin, G. Hays, A. H. Reid, D. Rupp, S. De Jong, S. Lee, W. S. Lee, Y. D. Chuang, R. Coffee, J. P. Cryan, J. M. Glownia, A. Föhlisch, M. R. Holmes, A. R. Fry, W. E. White, C. Bostedt, A. O. Scherz, H. A. Durr, and W. F. Schlotter, *X-Ray Pulse Preserving Single-Shot Optical Cross-Correlation Method for Improved Experimental Temporal Resolution*, Appl. Phys. Lett. **100**, 1 (2012).

[10] S. Schorb, T. Gorkhover, J. P. Cryan, J. M. Glownia, M. R. Bionta, R. N. Coffee, B. Erk, R. Boll, C. Schmidt, D. Rolles, A. Rudenko, A. Rouzee, M. Swiggers, S. Carron, J.-C. Castagna, J. D. Bozek, M. Messerschmidt, W. F. Schlotter, and C. Bostedt, *X-Ray–Optical Cross-Correlator for Gas-Phase Experiments at the Linac Coherent Light Source Free-Electron Laser*, Appl. Phys. Lett. **100**, 121107 (2012).

[11] H. C. Kapteyn, M. M. Murnane, and I. P. Christov, *Extreme Nonlinear Optics: Coherent X Rays from Lasers*, Phys. Today **58**, 39 (2005).

[12] C. Winterfeldt, C. Spielmann, and G. Gerber, *Colloquium: Optimal Control of High-Harmonic Generation*, Rev. Mod. Phys. **80**, 117 (2008).

[13] T. Lee, Y. Jiang, C. G. Rose-Petruck, and F. Benesch, *Ultrafast Tabletop Laser-Pump-x-Ray Probe Measurement of Solvated Fe(CN) 64-*, J. Chem. Phys. **122**, (2005).

[14] L. Miaja-Avila, G. C. O'Neil, J. Uhlig, C. L. Cromer, M. L. Dowell, R. Jimenez, A. S. Hoover,

K. L. Silverman, and J. N. Ullom, *Laser Plasma X-Ray Source for Ultrafast Time-Resolved x-Ray Absorption Spectroscopy*, Struct. Dyn. **2**, (2015).

[15] N. Zhavoronkov, Y. Gritsai, M. Bargheer, M. Woerner, T. Elsaesser, F. Zamponi, I. Uschmann, and E. Forster, *Ultrafast Optics - Microfocus Cu Ka Source for Femtosecond x-Ray Science*, Opt. Lett. **30**, 1737 (2005).

[16] S. Khan, K. Holldack, T. Kachel, R. Mitzner, and T. Quast, *Femtosecond Undulator Radiation from Sliced Electron Bunches*, Phys. Rev. Lett. **97**, 1 (2006).

[17] R. W. Schoenlein, S. Chattopadhyay, H. H. W. Chong, T. E. Glover, P. A. Heimann, C. V. Shank, A. A. Zholents, and M. S. Zolotorev, *Generation of Femtosecond Pulses of Synchrotron Radiation*, Science (80-. ). **287**, 2237 (2000).

[18] K. Yao, F. Willems, C. Von Korff Schmising, I. Radu, C. Strüber, D. Schick, D. Engel, A. Tsukamoto, J. K. Dewhurst, S. Sharma, and S. Eisebitt, *Distinct Spectral Response in $M$ -Edge Magnetic Circular Dichroism*, Phys. Rev. B **102**, 100405 (2020).

[19] M. Hennes, B. Rösner, V. Chardonnet, G. S. Chiuzbaian, R. Delaunay, F. Döring, V. A. Guzenko, M. Hehn, R. Jarrier, A. Kleibert, M. Lebugle, J. Lüning, G. Malinowski, A. Merhe, D. Naumenko, I. P. Nikolov, I. Lopez-Quintas, E. Pedersoli, T. Savchenko, B. Watts, M. Zangrando, C. David, F. Capotondi, B. Vodungbo, and E. Jal, *Time-Resolved XUV Absorption Spectroscopy and Magnetic Circular Dichroism at the Ni M2,3-Edges*, Appl. Sci. **11**, 325 (2020).

[20] S. Jana, R. S. Malik, Y. O. Kvashnin, I. L. M. Locht, R. Knut, R. Stefanuik, I. Di Marco, A. N. Yaresko, M. Ahlberg, J. Åkerman, R. Chimata, M. Battiato, J. Söderström, O. Eriksson, and O. Karis, *Analysis of the Linear Relationship between Asymmetry and Magnetic Moment at the $M$ Edge of $3d$ Transition Metals*, Phys. Rev. Res. **2**, 013180 (2020).

[21] S. Mathias, C. La-O-Vorakiat, P. Grychtol, P. Granitzka, E. Turgut, J. M. Shaw, R. Adam, H. T. Nembach, M. E. Siemens, S. Eich, C. M. Schneider, T. J. Silva, M. Aeschlimann, M. M. Murnane, and H. C. Kapteyn, *Probing the Timescale of the Exchange Interaction in a Ferromagnetic Alloy*, Proc. Natl. Acad. Sci. U. S. A. **109**, 4792 (2012).

[22] S. Jana, J. A. Terschlüsen, R. Stefanuik, S. Plogmaker, S. Troisi, R. S. Malik, M. Svanqvist, R. Knut, J. Söderström, and O. Karis, *A Setup for Element Specific Magnetization Dynamics Using the Transverse Magneto-Optic Kerr Effect in the Energy Range of 30-72 EV*, Rev. Sci. Instrum. **88**, 033113 (2017).

[23] C. Gahl, A. Azima, M. Beye, M. Deppe, K. Döbrich, U. Hasslinger, F. Hennies, A. Melnikov, M. Nagasono, A. Pietzsch, M. Wolf, W. Wurth, and A. Föhlisch, *A Femtosecond X-Ray/Optical Cross-Correlator*, Nat. Photonics **2**, 165 (2008).

[24] S. Schorb, T. Gorkhover, J. P. Cryan, J. M. Glownia, M. R. Bionta, R. N. Coffee, B. Erk, R. Boll, C. Schmidt, D. Rolles, A. Rudenko, A. Rouzee, M. Swiggers, S. Carron, J. C. Castagna, J. D. Bozek, M. Messerschmidt, W. F. Schlotter, and C. Bostedt, *X-Ray-Optical Cross-Correlator for Gas-Phase Experiments at the Linac Coherent Light Source Free-Electron Laser*, Appl. Phys. Lett. **100**, (2012).

[25] R. Riedel, A. Al-Shemmary, M. Gensch, T. Golz, M. Harmand, N. Medvedev, M. J. Prandolini, K. Sokolowski-Tinten, S. Toleikis, U. Wegner, B. Ziaja, N. Stojanovic, and F. Tavella, *Single-Shot Pulse Duration Monitor for Extreme Ultraviolet and X-Ray Free-Electron Lasers*, Nat. Commun. **4**, 1 (2013).

[26] M. Harmand, R. Coffee, M. R. Bionta, M. Chollet, D. French, D. Zhu, D. M. Fritz, H. T. Lemke, N. Medvedev, B. Ziaja, S. Toleikis, and M. Cammarata, *Achieving Few-Femtosecond Time-Sorting at Hard X-Ray Free-Electron Lasers*, Nat. Photonics **7**, 215 (2013).

[27] N. Hartmann, W. Helml, A. Galler, M. R. Bionta, J. Grünert, S. L. Molodtsov, K. R. Ferguson, S. Schorb, M. L. Swiggers, S. Carron, C. Bostedt, J. C. Castagna, J. Bozek, J. M. Glownia, D. J. Kane, A. R. Fry, W. E. White, C. P. Hauri, T. Feurer, and R. N. Coffee, *Sub-Femtosecond Precision Measurement of Relative X-Ray Arrival Time for Free-Electron Lasers*, Nat. Photonics **8**, 706 (2014).

[28] S. W. Epp, M. Hada, Y. Zhong, Y. Kumagai, K. Motomura, S. Mizote, T. Ono, S. Owada, D. Axford, S. Bakhtiarzadeh, H. Fukuzawa, Y. Hayashi, T. Katayama, A. Marx, H. M. Müller-Werkmeister, R. L. Owen, D. A. Sherrell, K. Tono, K. Ueda, F. Westermeier, and R. J. D. Miller, *Time Zero Determination for FEL Pump-Probe Studies Based on Ultrafast Melting of Bismuth*, Struct. Dyn. **4**, 054308 (2017).

[29] D. Schick, S. Eckert, N. Pontius, R. Mitzner, A. Föhlisch, K. Holldack, and F. Sorgenfrei, *Versatile Soft X-Ray-Optical Cross-Correlator for Ultrafast Applications*, Struct. Dyn. **3**, 054304 (2016).

[30] C. von Korff Schmising, M. Bargheer, M. Kiel, N. Zhavoronkov, M. Woerner, T. Elsaesser, I. Vrejoiu, D. Hesse, and M. Alexe, *Accurate Time Delay Determination for Femtosecond X-Ray Diffraction Experiments*, Appl. Phys. B **88**, 1 (2007).

[31] M. Bargheer, N. Zhavoronkov, Y. Gritsai, J. C. Woo, D. S. Kim, M. Woerner, and T. Elsaesser, *Coherent Atomic Motions in a Nanostructure Studied by Femtosecond X-Ray Diffraction*, Science (80-. ). **306**, 1771 (2004).

[32] T. Kachel, N. Pontius, C. Stamm, M. Wietstruk, E. F. Aziz, H. A. Dürr, W. Eberhardt, and F. M. F. de Groot, *Transient Electronic and Magnetic Structures of Nickel Heated by Ultrafast Laser Pulses*, Phys. Rev. B **80**, 092404 (2009).

[33] F. de Groot and A. Kotani, *Core Level Spectroscopy of Solids*, 1st ed. (CRC Press, Boca Raton, Florida, 2008).

[34] M. Volkov, S. A. Sato, F. Schlaepfer, L. Kasmi, N. Hartmann, M. Lucchini, L. Gallmann, A. Rubio, and U. Keller, *Attosecond Screening Dynamics Mediated by Electron Localization in Transition Metals*, Nat. Phys. **15**, 1145 (2019).

[35] H.-T. Chang, A. Guggenmos, S. K. Cushing, Y. Cui, N. U. Din, S. R. Acharya, I. J. Porter, U. Kleineberg, V. Turkowski, T. S. Rahman, D. M. Neumark, and S. R. Leone, *Electron Thermalization and Relaxation in Laser-Heated Nickel by Few-Femtosecond Core-Level Transient Absorption Spectroscopy*, Phys. Rev. B **103**, 64305 (2021).

[36] C. Stamm, T. Kachel, N. Pontius, R. Mitzner, T. Quast, K. Holldack, S. Khan, C. Lupulescu, E. F. Aziz, M. Wietstruk, H. A. A. Dürr, and W. Eberhardt, *Femtosecond Modification of Electron Localization and Transfer of Angular Momentum in Nickel*, Nat. Mater. **6**, 740 (2007).

[37] C. Stamm, N. Pontius, T. Kachel, M. Wietstruk, and H. A. Dürr, *Femtosecond X-Ray Absorption Spectroscopy of Spin and Orbital Angular Momentum in Photoexcited Ni Films during Ultrafast Demagnetization*, Phys. Rev. B - Condens. Matter Mater. Phys. **81**, 1 (2010).

[38] N. Rothenbach, M. E. Gruner, K. Ollefs, C. Schmitz-Antoniak, S. Salamon, P. Zhou, R. Li, M. Mo, S. Park, X. Shen, S. Weathersby, J. Yang, X. J. Wang, R. Pentcheva, H. Wende, U. Bovensiepen, K. Sokolowski-Tinten, and A. Eschenlohr, *Microscopic Nonequilibrium Energy Transfer Dynamics in a Photoexcited Metal/Insulator Heterostructure*, Phys. Rev. B **100**, 1 (2019).

[39] D. J. Higley, A. H. Reid, Z. Chen, L. Le Guyader, O. Hellwig, A. A. Lutman, T. Liu, P. Shafer, T. Chase, G. L. Dakovski, A. Mitra, E. Yuan, J. Schlappa, H. A. Dürr, W. F. Schlotter, and J. Stöhr, *Femtosecond X-Ray Induced Changes of the Electronic and Magnetic Response of Solids from Electron Redistribution*, Nat. Commun. **10**, 1 (2019).

[40] K. Carva, D. Legut, and P. M. Oppeneer, *Influence of Laser-Excited Electron Distributions on*

*the X-Ray Magnetic Circular Dichroism Spectra: Implications for Femtosecond Demagnetization in Ni*, Epl **86**, (2009).

[41] H.-S. Rhie, H. A. Dürr, and W. Eberhardt, *Femtosecond Electron and Spin Dynamics in Ni/W(110) Films*, Phys. Rev. Lett. **90**, 247201 (2003).

[42] V. P. Zhukov, E. V. Chulkov, and P. M. Echenique, *Lifetimes of Excited Electrons In Fe And Ni: First-Principles GW and the T -Matrix Theory*, Phys. Rev. Lett. **93**, 096401 (2004).

[43] V. P. Zhukov, E. V. Chulkov, and P. M. Echenique, *Lifetimes and Inelastic Mean Free Path of Low-Energy Excited Electrons in Fe, Ni, Pt, and Au: Ab Initio GW+T Calculations*, Phys. Rev. B - Condens. Matter Mater. Phys. **73**, 1 (2006).

[44] K. Holldack, J. Bahrdt, A. Balzer, U. Bovensiepen, M. Brzhezinskaya, A. Erko, A. Eschenlohr, R. Follath, A. Firsov, W. Frentrup, L. Le Guyader, T. Kachel, P. Kuske, R. Mitzner, R. Müller, N. Pontius, T. Quast, I. Radu, J.-S. Schmidt, C. Schüßler-Langeheine, M. Sperling, C. Stamm, C. Trabant, and A. Föhlisch, *FemtoSpeX: A Versatile Optical Pump–Soft X-Ray Probe Facility with 100 Fs X-Ray Pulses of Variable Polarization*, J. Synchrotron Radiat. **21**, 1090 (2014).

[45] A. Bojahr, D. Schick, L. Maerten, M. Herzog, I. Vrejoiu, C. von Korff Schmising, C. Milne, S. L. Johnson, and M. Bargheer, *Comparing the Oscillation Phase in Optical Pump-Probe Spectra to Ultrafast x-Ray Diffraction in the Metal-Dielectric SrRuO$_3$/SrTiO$_3$<*, Phys. Rev. B **85**, 224302 (2012).